\documentclass[twocolumn,tighten,dvipsnames]{aastex631}

\usepackage{amsmath}
\usepackage{xspace}

\usepackage{booktabs}
\usepackage{xcolor}

\usepackage[ruled]{algorithm2e}

\graphicspath{{figs}}

\newcommand{\PADC}{$P$-ADC\xspace}
\newcommand{\ADC}{\mathrm{ADC}}
\newcommand{\nmodP}{{n\,\mathrm{mod}\,P}}
\newcommand{\CC}{\mbox{C.C.}}

\newcommand{\RMS}{\mathrm{RMS}}
\newcommand{\LCM}{\mathrm{LCM}}

\begin{document}

\title{A Direct Calibration Algorithm for ADC Interleaving}

\author[0000-0001-6337-6126]{Chi-kwan Chan}
\affiliation{Steward Observatory and Department of Astronomy, University of Arizona, 933 N. Cherry Ave., Tucson, AZ~85721, USA}
\affiliation{Data Science Institute, University of Arizona, 1230 N. Cherry Ave., Tucson, AZ~85721, USA}
\affiliation{Program in Applied Mathematics, University of Arizona, 617 N. Santa Rita, Tucson, AZ~85721, USA}

\author[0009-0003-3707-5938]{Hina Suzuki}
\affiliation{Steward Observatory and Department of Astronomy, University of Arizona, 933 N. Cherry Ave., Tucson, AZ~85721, USA}
\affiliation{Department of Electrical and Computer Engineering, University of Arizona, 1230 E. Speedway Blvd., Tucson, AZ~85721, USA}

\author{David Forbes}
\affiliation{Steward Observatory and Department of Astronomy, University of Arizona, 933 N. Cherry Ave., Tucson, AZ~85721, USA}

\author{Andrew Thomas West}
\affiliation{Steward Observatory and Department of Astronomy, University of Arizona, 933 N. Cherry Ave., Tucson, AZ~85721, USA}

\author{Arash Roshanineshat}
\affiliation{Texas Instruments Inc, 251 S Williams Blvd, Tucson, AZ~85711, USA}
\affiliation{Steward Observatory and Department of Astronomy, University of Arizona, 933 N. Cherry Ave., Tucson, AZ~85721, USA}
\affiliation{Department of Electrical and Computer Engineering, University of Arizona, 1230 E. Speedway Blvd., Tucson, AZ~85721, USA}

\author[0000-0002-2367-1080]{Daniel P. Marrone}
\affiliation{Steward Observatory and Department of Astronomy, University of Arizona, 933 N. Cherry Ave., Tucson, AZ~85721, USA}

\author[0000-0002-4747-4276]{Amy Lowitz}
\affiliation{Steward Observatory and Department of Astronomy, University of Arizona, 933 N. Cherry Ave., Tucson, AZ~85721, USA}

\begin{abstract}
  We introduce a novel direct calibration algorithm to address phase
  delay, gain, and offset mismatches in Analog-to-Digital Converter
  (ADC) time interleaving systems.
  These mismatches, common in high-speed data acquisition, degrade
  system performance and signal integrity, particularly in
  applications such as radio astronomy and very long baseline
  interferometry (VLBI).
  Our proposed algorithm uses a sinusoidal reference signal and
  Fourier analysis to isolate and correct each type of mismatch,
  providing a computationally efficient solution.
  Extensive numerical simulations validate the algorithm's
  effectiveness and demonstrate its ability to significantly enhance
  signal reconstruction accuracy compared to existing methods.
  This work provides a robust and scalable solution for maintaining
  signal fidelity in interleaved ADC systems and has broad
  applications in fields that require high-speed data acquisition.
\end{abstract}


\section{Introduction}
\label{sec:intro}

The accurate conversion of analog signals to digital is fundamental
for a wide range of scientific and engineering applications, including
radio astronomy~\citep{2003ASPC..306..161I, 2009IEEEP..97.1482D,
  2019ApJ...875L...2E, 2019ApJ...875L...3E, 2020PASP..132h5001J,
  2022ApJ...930L..13E}, telecommunications, medical imaging, and
various other scientific and industrial fields.
This conversion is carried out by Analog-to-Digital Converters (ADCs),
which sample continuous-time signals and convert them into
discrete-time digital values.
To achieve high sampling rates and enhance time resolution, many
modern systems~\citep[e.g.,][]{Fujitsu2010, 2014JAI.....350001P}
employ ADC time interleaving---a technique in which multiple ADCs
operate in parallel, each sampling the input signal at staggered
intervals.
By combining the outputs of these interleaved ADCs, the effective
sampling rate is increased, thereby enhancing the overall performance
of the system.

Despite its advantages, ADC interleaving introduces challenges due to
mismatches among the interleaved ADCs.
These mismatches can distort the signal and introduce spurious tones
in the reconstructed signal~\citep{Elbornsson2005, Roshanineshat2023}.
In VLBI systems, such as the Event Horizon Telescope (EHT)'s digital
backend, these interleaving mismatches have been shown to produce
persistent spectral artifacts that degrade signal fidelity and
interfere with the detection of weak astronomical signals~\citep[see,
  e.g.,][]{2014JAI.....350001P}.
Addressing these mismatches is critical for maintaining signal
integrity and achieving the desired system performance.

Several calibration techniques have been proposed to mitigate the
effects of these mismatches~\citep[e.g.,][]{Reeder2005, Manganaro2015,
  Shen2022, Sjoland2022, Hu2022, Sundstrom2023}.
While these methods have demonstrated success in correcting certain
types of mismatches, many systems still face challenges in achieving
robust, high-fidelity performance across all mismatch types.

In this paper, we introduce a novel direct calibration algorithm
designed to correct phase delay, gain, and offset mismatches in ADC
interleaving systems.
Our approach leverages the elegant properties of the Fourier transform
to simultaneously isolate and solve for each type of mismatch.
This process is computationally efficient and highly effective, making
it suitable for high-speed signal processing applications such as
VLBI, where real-time calibration is crucial to preserving signal
integrity over extended observation periods.

The paper is organized as follows.
In section~\ref{sec:setup}, we define the setup and notations used
throughout the paper.
In section~\ref{sec:math}, we derives the mathematical identities that
form the foundation of our direct calibration algorithm.
We describe the minimal sampling requirement and explore the
algorithm's convergence properties in sections~\ref{sec:limit} and
\ref{sec:noise}.
In section~\ref{sec:algo}, we present the pseudocode that can be
directly used in practical systems.
Finally, we conclude the paper in section~\ref{sec:discussions},
summarizing our contributions and discussing their potential
applications.

\section{Setup and Notations}
\label{sec:setup}

\begin{figure}[b]
  \centering
  \includegraphics[width=\columnwidth]{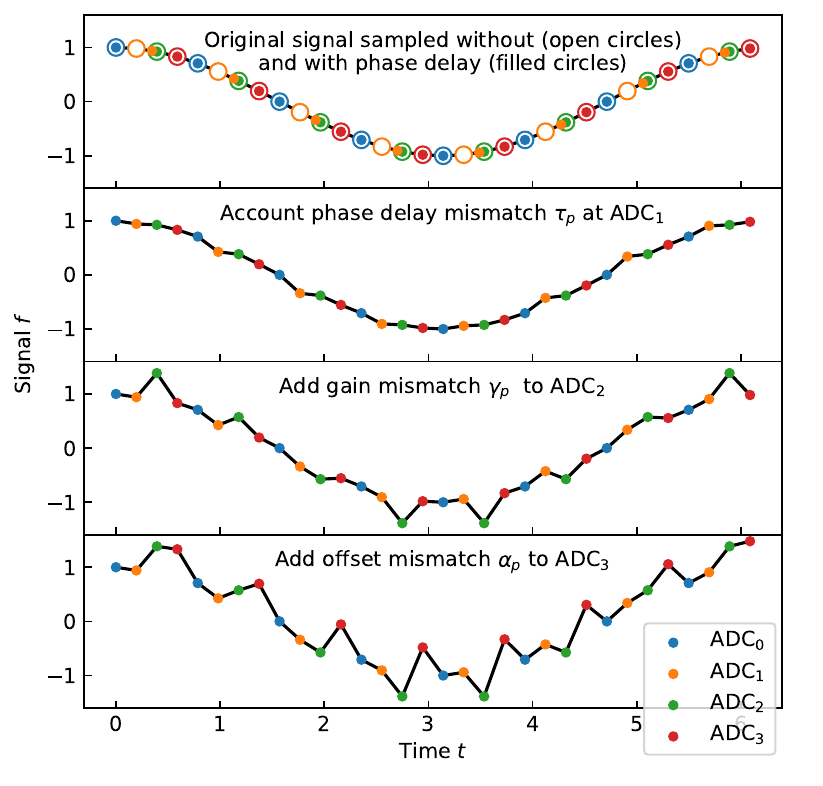}
  \caption{Illustration of the impact of mismatches in a 4-ADC
    interleaving system.
    The black solid curve in the shows the input signal.
    Open circles mark the ideal sampling points for a mismatch-free
    system, while filled circles show the actual sampling locations
    when mismatches are present.
    The top panel shows the ideal case so the black curve is
    $\cos(t)$.
    We introduce a phase delay mismatch, which samples the signal in a
    non-uniform fashion.
    The second panel highlights the effect of a phase delay mismatch
    in $\ADC_1$, which results in a distortion in the reconstructed
    signal.
    The third panel adds a gain mismatch in $\ADC_2$, further
    modifying the amplitude of the sampled signal.
    The fourth panel introduces an offset mismatch in $\ADC_3$,
    demonstrating how combined mismatches lead to significant distortion.
    Accurate calibration of these mismatches is essential for
    high-fidelity signal reconstruction in interleaved ADC systems.}
  \label{fig:mismatches}
\end{figure}

We consider a generic setup involving $P$ identical ADCs, which we
label as $\ADC_0$, $\ADC_1$, \dots, $\ADC_{P-1}$.
Each ADC has a sampling interval of $\Delta$, resulting in an
individual sampling rate of $1/\Delta$.
To achieve a higher effective sampling rate, we stagger the sampling
times of these ADCs.
Specifically, $\ADC_p$ samples the input signal with an additional
temporal offset of $p\,\Delta/P = p\,\delta$, where $\delta \equiv
\Delta/P$.

By interleaving the ADCs, the system samples the input signal at the
following times:
\begin{align}
  t_{p,n'} \equiv n' \Delta + p\,\delta.
\end{align}
Sorting these sampling times in ascending order, the overall system
samples at
\begin{align}
  t_n \equiv n\,\delta.
\end{align}
In the above definitions, both $n$ and $n'$ are integers.
It is also clear that the sample taken at time $t_n$ originates from
$\ADC_\nmodP$.
We assume that the ``integration time'' or aperture width of each ADC
is much shorter than $\delta$, so that each sample can be treated as
an instantaneous value.

Practical implementations of ADC interleaving often encounter
mismatches between the ADCs, which can significantly degrade system
performance.
Following \citet{Roshanineshat2023}, we ignore bandwidth mismatches
and focus on three primary types of mismatches:
\emph{i})~phase delay mismatches $\tau_p$,
\emph{ii})~gain mismatches $\gamma_p$, and
\emph{iii})~offset mismatches $\alpha_p$.
Phase delay mismatches refer to timing deviations in the effective
sampling instants of the ADCs, typically caused by differences in
clock routing or circuitry delays across channels.
Gain mismatches represent variations in the amplification factors of
the ADCs, and offset mismatches denote differences in their baseline
DC levels.
Both gain and offset mismatches usually arise from manufacturing
variations between individual ADCs.
The combined effect of these mismatches is illustrated in
Figure~\ref{fig:mismatches}.

To model the sampled data from $\ADC_p$ in the presence of these
mismatches, we use the following expression:
\begin{align}
  \tilde{f}_{p,n'} = \gamma_p f(n'\Delta + p\,\delta + \tau_p) + \alpha_p,
\end{align}
where the tilde ($\tilde{\ }$) denotes that the sampled data includes
mismatch effects.
Rewriting this in terms of the global interleaved sampling index $n$,
we obtain:
\begin{align}
  \tilde{f}_n = \gamma_\nmodP f(n\,\delta + \tau_\nmodP) + \alpha_\nmodP.
  \label{eq:mismatches}
\end{align}

To calibrate the system and correct these mismatches, we use a
sinusoidal reference signal of known period $T$:
\begin{align}
  f(t) = A \cos(\Omega\,t) = \frac{A}{2}\left[e^{i\Omega\,t} + \CC\right],
\end{align}
where $\Omega \equiv 2\pi/T$ and ``$\CC$'' denotes the complex
conjugate of the preceding term.
The sampled/discretized version of this signal, affected by the
mismatches in equation~(\ref{eq:mismatches}), is therefore given by
\begin{align}
  \tilde{f}_n
  = \gamma_\nmodP\frac{A}{2}
  \left[e^{i\Omega(n\delta + \tau_\nmodP)} + \CC\right]
  + \alpha_\nmodP.
  \label{eq:sampled}
\end{align}

Throughout this paper, we define the discrete-time Fourier transform
of the sampled data $\tilde{f}_n$ as
\begin{align}
  \tilde{\mathcal{F}}_\omega
  = \sum_{n=-\infty}^\infty e^{-i\omega n\delta} \tilde{f}_n.
  \label{eq:FT}
\end{align}
This formulation provides the basis for our calibration algorithm
described in the following sections.

\section{Mathematical Derivation}
\label{sec:math}

In this section, we derive the mathematical identities that form the
foundation of our direct calibration algorithm to correct phase delay,
gain, and offset mismatches.
The key idea is to analyze the Fourier coefficients of the sampled
sinusoidal reference signal, which encode the mismatch parameters in a
structured way.

\begin{widetext}

Substituting the sampled signal with mismatches from
equation~(\ref{eq:sampled}) into the definition of the Fourier
transform in equation~(\ref{eq:FT}), we obtain
\begin{align}
  \tilde{\mathcal{F}}_\omega
  &= \sum_{n=-\infty}^\infty e^{-i\omega n\delta} \left\{
       \gamma_\nmodP\frac{A}{2}\left[e^{i\Omega(n\delta + \tau_\nmodP)} + \CC\right] + \alpha_\nmodP
     \right\} \\
  &= \sum_{n'=-\infty}^\infty \sum_{p=0}^{P-1} e^{-i\omega(n'P + p)\delta} \left\{
       \gamma_p\frac{A}{2}\left[e^{i\Omega[(n'P + p)\delta + \tau_p]} + \CC\right] + \alpha_p
     \right\} \\
  &= \frac{A}{2}
     \sum_{n'=-\infty}^\infty e^{-i\omega n'P\delta + i\Omega n'P\delta}
     \sum_{p=0}^{P-1} \gamma_p \left[e^{-i\omega p\delta + i\Omega(p\delta + \tau_p)} + \CC\right]
   + \sum_{n'=-\infty}^\infty e^{-i\omega n'P\delta} \sum_{p=0}^{P-1} e^{-i\omega p\delta}\alpha_p \\
  &= \frac{A}{2}
     \sum_{n'=-\infty}^\infty e^{-i(\omega - \Omega)n'\Delta}
     \sum_{p=0}^{P-1} \gamma_p \left[e^{-i(\omega - \Omega)p\delta + i\Omega\tau_p} + \CC\right]
   + \sum_{n'=-\infty}^\infty e^{-i\omega n' \Delta} \sum_{p=0}^{P-1} e^{-i\omega p\delta}\alpha_p
  \label{eq:F}
\end{align}

The two infinite sums over $n'$ in equation~(\ref{eq:F}) are non-zero
only if $\Delta(\omega - \Omega) = 2\pi m$ and $\Delta \omega = 2\pi
m'$, respectively, for some integers $m$ and $m'$.
Letting $\Omega_\Delta \equiv 2\pi/\Delta$, equation~(\ref{eq:F})
simplifies to
\begin{align}
  \tilde{\mathcal{F}}_\omega
  &= \frac{A}{2}\delta(\omega - \Omega - m\Omega_\Delta)
     \sum_{p=0}^{P-1} \gamma_p \left[e^{-i(\omega - \Omega)p\delta + i\Omega\tau_p} + \CC\right]
   + \delta(\omega - m'\Omega_\Delta)\sum_{p=0}^{P-1}e^{-i\omega p\delta}\alpha_p.
\end{align}
\emph{Assuming that $\Omega$ is not an integer multiple of
$\Omega_\Delta$}, the two types of nonzero Fourier coefficients are:
\begin{align}
  \tilde{\mathcal{F}}_{\Omega + m\Omega_\Delta}
  &= \frac{A}{2}\sum_{p=0}^{P-1} e^{-im\Omega_\Delta p\delta}
     \gamma_p e^{i\Omega\tau_p}
   = \frac{A}{2}\sum_{p=0}^{P-1} e^{-2\pi imp/P} \gamma_p e^{i\Omega\tau_p},
  \label{eq:classI} \\
  \tilde{\mathcal{F}}_{m'\Omega_\Delta}
  &= \sum_{p=0}^{P-1} e^{-im'\Omega_\Delta p\delta}\alpha_p
   = \sum_{p=0}^{P-1} e^{-2\pi im'p/P}\alpha_p.
  \label{eq:classII}
\end{align}

\end{widetext}

To derive our calibration algorithm, according to
equation~(\ref{eq:classI}), we define
\begin{align}
  \tilde{\mathcal{R}}_m \equiv
  \frac{\tilde{\mathcal{F}}_{\Omega + m\Omega_\Delta}}{A/2}
\end{align}
as the \emph{dimensionless ratio} of the non-zero coefficients and
$A/2$, and
\begin{align}
  \varGamma_p \equiv \gamma_p e^{i\Omega\tau_p}
\end{align}
as the \emph{complex gain} for $\ADC_p$ on the reference signal, $f(t)
= A\cos(\Omega t)$, which encapsulates both gain and phase delay
mismatches.
According to equation~(\ref{eq:classII}), it is convenient to
define
\begin{align}
  \tilde{\mathcal{A}}_m \equiv \tilde{\mathcal{F}}_{m\Omega_\Delta},
\end{align}
where we drop the prime for notational simplicity.
With these definitions, equations~(\ref{eq:classI}) and
(\ref{eq:classII}) reduce to
\begin{align}
  \tilde{\mathcal{R}}_{m}
  &= \sum_{p=0}^{P-1}e^{-2\pi im p/P} \varGamma_p,
  \label{eq:Rm} \\
  \tilde{\mathcal{A}}_{m}
  &= \sum_{p=0}^{P-1}e^{-2\pi im p/P} \alpha_p.
  \label{eq:Am}
\end{align}

Equations~\eqref{eq:Rm} and \eqref{eq:Am} are the main result of this
paper.
They show that $\tilde{\mathcal{R}}_m$ and $\tilde{\mathcal{A}}_m$ are
simply the discrete Fourier transforms (DFTs) of the complex gain
$\varGamma_p$ and the offset mismatch $\alpha_p$, respectively.
These identities generalize equation~(7) in \citet{Shen2022}.
They can also be seen as a special cases of a standard decomposition
commonly used in parallelizing one-dimensional fast Fourier
transforms~(FFTs).

\begin{table*}
  \setlength{\tabcolsep}{18pt}
  \renewcommand{\arraystretch}{0.75}
  \footnotesize
  \centering
  \begin{tabular}{|crr|c|cc|c|}
    \hline
    $P$ & $N_\delta$ & $N_T N_\delta$ & $N_T N_\delta/P$ & $\RMS(\hat\gamma_p-\gamma_p)$ & $\RMS(\hat\tau_p-\tau_p)$ & $\RMS(\hat\alpha_p-\alpha_p)$ \\
    \hline
    2 &  2~ &  2~~~ & {\color{black}        1} & \texttt{\color{black}       1.896e+00} & \texttt{\color{black}       9.435e-04} & \texttt{\color{black}       8.916e-01} \\
    2 &  3~ &  6~~~ & {\color{MidnightBlue} 3} & \texttt{\color{MidnightBlue}3.237e-16} & \texttt{\color{MidnightBlue}3.019e-18} & \texttt{\color{MidnightBlue}2.614e-16} \\
    2 &  4~ &  4~~~ & {\color{OliveGreen}   2} & \texttt{\color{black}       7.472e-01} & \texttt{\color{black}       3.395e-03} & \texttt{\color{OliveGreen}  4.763e-17} \\
    2 &  5~ & 10~~~ & {\color{MidnightBlue} 5} & \texttt{\color{MidnightBlue}1.755e-16} & \texttt{\color{MidnightBlue}1.238e-18} & \texttt{\color{MidnightBlue}1.870e-16} \\
    2 &  6~ &  6~~~ & {\color{MidnightBlue} 3} & \texttt{\color{MidnightBlue}7.850e-17} & \texttt{\color{MidnightBlue}2.719e-18} & \texttt{\color{MidnightBlue}1.376e-16} \\
    2 &  7~ & 14~~~ & {\color{MidnightBlue} 7} & \texttt{\color{MidnightBlue}1.755e-16} & \texttt{\color{MidnightBlue}2.892e-18} & \texttt{\color{MidnightBlue}6.090e-17} \\
    2 &  8~ &  8~~~ & {\color{MidnightBlue} 4} & \texttt{\color{MidnightBlue}0.000e+00} & \texttt{\color{MidnightBlue}1.333e-18} & \texttt{\color{MidnightBlue}2.314e-17} \\
    2 &  9~ & 18~~~ & {\color{MidnightBlue} 9} & \texttt{\color{MidnightBlue}1.570e-16} & \texttt{\color{MidnightBlue}3.803e-18} & \texttt{\color{MidnightBlue}1.972e-16} \\
    2 & 10~ & 10~~~ & {\color{MidnightBlue} 5} & \texttt{\color{MidnightBlue}1.755e-16} & \texttt{\color{MidnightBlue}8.606e-19} & \texttt{\color{MidnightBlue}3.035e-17} \\
    2 & 11~ & 22~~~ & {\color{MidnightBlue}11} & \texttt{\color{MidnightBlue}2.220e-16} & \texttt{\color{MidnightBlue}1.516e-18} & \texttt{\color{MidnightBlue}1.097e-17} \\
    3 &  2~ &  6~~~ & {\color{OliveGreen}   2} & \texttt{\color{black}       8.617e-01} & \texttt{\color{black}       8.668e-04} & \texttt{\color{OliveGreen}  9.363e-17} \\
    3 &  3~ &  3~~~ & {\color{black}        1} & \texttt{\color{black}       1.431e+00} & \texttt{\color{black}       3.351e-03} & \texttt{\color{black}       8.124e-01} \\
    3 &  4~ & 12~~~ & {\color{MidnightBlue} 4} & \texttt{\color{MidnightBlue}6.410e-17} & \texttt{\color{MidnightBlue}5.056e-18} & \texttt{\color{MidnightBlue}4.446e-16} \\
    3 &  5~ & 15~~~ & {\color{MidnightBlue} 5} & \texttt{\color{MidnightBlue}5.477e-16} & \texttt{\color{MidnightBlue}4.190e-18} & \texttt{\color{MidnightBlue}4.502e-16} \\
    3 &  6~ &  6~~~ & {\color{OliveGreen}   2} & \texttt{\color{black}       6.277e-01} & \texttt{\color{black}       5.652e-03} & \texttt{\color{OliveGreen}  1.540e-16} \\
    3 &  7~ & 21~~~ & {\color{MidnightBlue} 7} & \texttt{\color{MidnightBlue}1.433e-16} & \texttt{\color{MidnightBlue}5.274e-18} & \texttt{\color{MidnightBlue}1.152e-16} \\
    3 &  8~ & 24~~~ & {\color{MidnightBlue} 8} & \texttt{\color{MidnightBlue}1.923e-16} & \texttt{\color{MidnightBlue}5.054e-18} & \texttt{\color{MidnightBlue}2.873e-16} \\
    3 &  9~ &  9~~~ & {\color{MidnightBlue} 3} & \texttt{\color{MidnightBlue}2.311e-16} & \texttt{\color{MidnightBlue}2.731e-18} & \texttt{\color{MidnightBlue}1.622e-16} \\
    3 & 10~ & 30~~~ & {\color{MidnightBlue}10} & \texttt{\color{MidnightBlue}1.813e-16} & \texttt{\color{MidnightBlue}6.862e-18} & \texttt{\color{MidnightBlue}2.125e-16} \\
    3 & 11~ & 33~~~ & {\color{MidnightBlue}11} & \texttt{\color{MidnightBlue}1.282e-16} & \texttt{\color{MidnightBlue}2.783e-18} & \texttt{\color{MidnightBlue}1.156e-16} \\
    4 &  2~ &  4~~~ & {\color{black}        1} & \texttt{\color{black}       8.975e-01} & \texttt{\color{black}       7.465e-04} & \texttt{\color{black}       9.439e-01} \\
    4 &  3~ & 12~~~ & {\color{OliveGreen}   3} & \texttt{\color{black}       1.561e-02} & \texttt{\color{black}       5.026e-05} & \texttt{\color{OliveGreen}  2.263e-16} \\
    4 &  4~ &  4~~~ & {\color{black}        1} & \texttt{\color{black}       1.091e+00} & \texttt{\color{black}       5.208e-03} & \texttt{\color{black}       7.265e-01} \\
    4 &  5~ & 20~~~ & {\color{MidnightBlue} 5} & \texttt{\color{MidnightBlue}2.989e-16} & \texttt{\color{MidnightBlue}6.033e-18} & \texttt{\color{MidnightBlue}5.752e-16} \\
    4 &  6~ & 12~~~ & {\color{MidnightBlue} 3} & \texttt{\color{MidnightBlue}2.289e-16} & \texttt{\color{MidnightBlue}3.965e-18} & \texttt{\color{MidnightBlue}1.597e-16} \\
    4 &  7~ & 28~~~ & {\color{MidnightBlue} 7} & \texttt{\color{MidnightBlue}5.118e-16} & \texttt{\color{MidnightBlue}9.702e-18} & \texttt{\color{MidnightBlue}2.091e-16} \\
    4 &  8~ &  8~~~ & {\color{OliveGreen}   2} & \texttt{\color{black}       7.197e-01} & \texttt{\color{black}       7.778e-03} & \texttt{\color{OliveGreen}  4.284e-17} \\
    4 &  9~ & 36~~~ & {\color{MidnightBlue} 9} & \texttt{\color{MidnightBlue}2.989e-16} & \texttt{\color{MidnightBlue}7.416e-18} & \texttt{\color{MidnightBlue}1.901e-16} \\
    4 & 10~ & 20~~~ & {\color{MidnightBlue} 5} & \texttt{\color{MidnightBlue}2.776e-16} & \texttt{\color{MidnightBlue}8.162e-18} & \texttt{\color{MidnightBlue}1.516e-16} \\
    4 & 11~ & 44~~~ & {\color{MidnightBlue}11} & \texttt{\color{MidnightBlue}2.719e-16} & \texttt{\color{MidnightBlue}3.811e-18} & \texttt{\color{MidnightBlue}2.730e-16} \\
    5 &  2~ & 10~~~ & {\color{OliveGreen}   2} & \texttt{\color{black}       8.991e-01} & \texttt{\color{black}       7.319e-04} & \texttt{\color{OliveGreen}  2.340e-16} \\
    5 &  3~ & 15~~~ & {\color{OliveGreen}   3} & \texttt{\color{black}       1.300e-01} & \texttt{\color{black}       3.390e-04} & \texttt{\color{OliveGreen}  8.141e-16} \\
    5 &  4~ & 20~~~ & {\color{OliveGreen}   4} & \texttt{\color{black}       4.699e-02} & \texttt{\color{black}       1.841e-04} & \texttt{\color{OliveGreen}  6.474e-16} \\
    5 &  5~ &  5~~~ & {\color{black}        1} & \texttt{\color{black}       1.005e+00} & \texttt{\color{black}       4.815e-03} & \texttt{\color{black}       7.099e-01} \\
    5 &  6~ & 30~~~ & {\color{MidnightBlue} 6} & \texttt{\color{MidnightBlue}4.578e-16} & \texttt{\color{MidnightBlue}8.607e-18} & \texttt{\color{MidnightBlue}8.437e-16} \\
    5 &  7~ & 35~~~ & {\color{MidnightBlue} 7} & \texttt{\color{MidnightBlue}5.371e-16} & \texttt{\color{MidnightBlue}1.520e-17} & \texttt{\color{MidnightBlue}4.529e-16} \\
    5 &  8~ & 40~~~ & {\color{MidnightBlue} 8} & \texttt{\color{MidnightBlue}3.179e-16} & \texttt{\color{MidnightBlue}1.006e-17} & \texttt{\color{MidnightBlue}2.153e-16} \\
    5 &  9~ & 45~~~ & {\color{MidnightBlue} 9} & \texttt{\color{MidnightBlue}5.088e-16} & \texttt{\color{MidnightBlue}1.592e-17} & \texttt{\color{MidnightBlue}3.906e-16} \\
    5 & 10~ & 10~~~ & {\color{OliveGreen}   2} & \texttt{\color{black}       6.321e-01} & \texttt{\color{black}       8.801e-03} & \texttt{\color{OliveGreen}  8.631e-17} \\
    5 & 11~ & 55~~~ & {\color{MidnightBlue}11} & \texttt{\color{MidnightBlue}2.809e-16} & \texttt{\color{MidnightBlue}1.028e-17} & \texttt{\color{MidnightBlue}1.288e-16} \\
    6 &  2~ &  6~~~ & {\color{black}        1} & \texttt{\color{black}       8.961e-01} & \texttt{\color{black}       6.987e-04} & \texttt{\color{black}       9.520e-01} \\
    6 &  3~ &  6~~~ & {\color{black}        1} & \texttt{\color{black}       7.023e-01} & \texttt{\color{black}       2.004e-03} & \texttt{\color{black}       7.973e-01} \\
    6 &  4~ & 12~~~ & {\color{OliveGreen}   2} & \texttt{\color{black}       8.840e-01} & \texttt{\color{black}       3.830e-03} & \texttt{\color{OliveGreen}  2.891e-16} \\
    6 &  5~ & 30~~~ & {\color{OliveGreen}   5} & \texttt{\color{black}       1.080e-02} & \texttt{\color{black}       4.614e-05} & \texttt{\color{OliveGreen}  6.736e-16} \\
    6 &  6~ &  6~~~ & {\color{black}        1} & \texttt{\color{black}       1.032e+00} & \texttt{\color{black}       8.547e-03} & \texttt{\color{black}       6.682e-01} \\
    6 &  7~ & 42~~~ & {\color{MidnightBlue} 7} & \texttt{\color{MidnightBlue}6.362e-16} & \texttt{\color{MidnightBlue}1.458e-17} & \texttt{\color{MidnightBlue}6.956e-16} \\
    6 &  8~ & 24~~~ & {\color{MidnightBlue} 4} & \texttt{\color{MidnightBlue}3.654e-16} & \texttt{\color{MidnightBlue}4.031e-18} & \texttt{\color{MidnightBlue}2.304e-16} \\
    6 &  9~ & 18~~~ & {\color{MidnightBlue} 3} & \texttt{\color{MidnightBlue}2.311e-16} & \texttt{\color{MidnightBlue}6.628e-18} & \texttt{\color{MidnightBlue}2.286e-16} \\
    6 & 10~ & 30~~~ & {\color{MidnightBlue} 5} & \texttt{\color{MidnightBlue}2.794e-16} & \texttt{\color{MidnightBlue}7.599e-18} & \texttt{\color{MidnightBlue}4.026e-16} \\
    6 & 11~ & 66~~~ & {\color{MidnightBlue}11} & \texttt{\color{MidnightBlue}7.166e-16} & \texttt{\color{MidnightBlue}1.052e-17} & \texttt{\color{MidnightBlue}2.740e-16} \\
    \hline
  \end{tabular}
  \caption{%
    Summary of a representative set of numerical experiments on
    recovering mismatch parameters in a $P$-ADC interleaving system.
    Here, $P$ is the number of interleaved ADCs; $N_\delta$ is the
    total number of samples per reference signal period across all
    ADCs; and $N_T N_\delta = \LCM(P, N_\delta)$ is the minimum total
    number of sampling points required so that each ADC receives an
    equal number of samples.
    Consequently, $N_T N_\delta / P$ is the number of samples acquired
    by each individual ADC.
    For each experiment, we randomly generate the ground-truth
    mismatch parameters $\hat\gamma_p$, $\hat\tau_p$, and
    $\hat\alpha_p$, then sample synthetic data and solve for the
    recovered values $\gamma_p$, $\tau_p$, and $\alpha_p$ using
    equations~\eqref{eq:Gamma_sol} and~\eqref{eq:alpha_sol}.
    We highlight the results in {\color{MidnightBlue}blue} when both
    conditions $N_\delta > P$ and $N_T N_\delta \ge 3P$ are satisfied,
    which meets the theoretical requirements for accurately recovering
    all mismatch parameters.
    Otherwise, if only $N_T N_\delta \ge 2P$ is satisfied, the results
    are shown in {\color{OliveGreen}green}, indicating configurations
    sufficient for recovering offset mismatches, but not necessarily
    gain or phase delay mismatches.%
  }
  \label{tab:errors}
\end{table*}

Since both $\tilde{\mathcal{R}}_m$ and $\tilde{\mathcal{A}}_m$ are
directly computable from the Fourier transform of the sampled data
$\tilde{f}_n$, we can recover $\varGamma_p$ and $\alpha_p$ by applying
the inverse DFT:
\begin{align}
  \varGamma_p &=
  \mbox{DFT}^{-1} \tilde{\mathcal{R}}_{m} \equiv
  \frac{1}{P} \sum_{p=0}^{P-1}e^{2\pi im p/P} \tilde{\mathcal{R}}_{m}\,,
  \label{eq:Gamma_sol}\\
  \alpha_p &=
  \mbox{DFT}^{-1} \tilde{\mathcal{A}}_{m} \equiv
  \frac{1}{P} \sum_{p=0}^{P-1}e^{2\pi im p/P} \tilde{\mathcal{A}}_{m}\,.
  \label{eq:alpha_sol}
\end{align}
The mismatches can then be solved by:
\begin{align}
\mbox{Gain mismatch: }        & \gamma_p = |\varGamma_p| \\
\mbox{Phase delay mismatch: } & \tau_p   = \arg(\varGamma_p)/\Omega \\
\mbox{Offset mismatch: }      & \alpha_p \mbox{ (computed directly)}
\end{align}
These quantities can be used to correct for mismatches either through
hardware adjustment or post-processing.
The ability to solve for all mismatch parameters in closed form makes
this calibration algorithm both efficient and scalable, especially in
high-speed interleaved systems with many ADC cores.

\section{Minimal Sampling Requirement}
\label{sec:limit}

Equations~\eqref{eq:Rm} and~\eqref{eq:Am} provide deterministic
relations between the measured Fourier coefficients and the mismatch
parameters.
In the absence of noise, and with a sufficient number of samples, all
mismatch parameters can be solved exactly, limited only by machine
precision.
This raises two interesting questions:
\emph{i})~What is the minimal number of sampling points per reference
signal period $T$ required to recover all mismatch parameters in a
\PADC interleaving system?
\emph{ii})~For a fixed number of sampling points per period
$N_\delta$, how many ADCs $P$ can be interleaved while still enabling
accurate calibration?

Throughout this section, we assume that the calibration signal
frequency $\Omega = 2\pi/T$ is \emph{not} an integer multiple of the
interleaved sampling frequency $\Omega_\Delta = 2\pi/\Delta$.
As discussed in the last section, this ensures that
$\tilde{\mathcal{R}}_m$ and $\tilde{\mathcal{A}}_m$ do not mix.
We begin by analyzing the recovery of offset mismatches $\alpha_p$.
To solve for each $\alpha_p$ individually, we must ensure that each
ADC samples the signal an equal number of times.
This occurs when the total number of samples is a multiple of the
least common multiple (LCM), i.e.,
\begin{align}
  N_T N_\delta = \LCM(P, N_\delta),
\end{align}
where $N_\delta$ is the number of samples taken per reference signal
period $T$ across the entire interleaved system, and $N_T$ is the
number of periods required to achieve balanced sampling across all $P$
ADCs.
As long as $N_T N_\delta \ge 2P$, each ADC contributes at least two
samples per period, which is sufficient to solve for the DC offset
values $\alpha_p$ to machine accuracy, even if the signal is not
Nyquist-sampled.

Recovering the gain $\gamma_p$ and phase delay $\tau_p$ mismatches
requires stricter conditions.
To capture meaningful phase information, the system must have more
samples per period than ADCs, i.e., $N_\delta > P$.
Additionally, we require at least $N_T N_\delta \ge 3P$ total samples
so that the system of equations has sufficient degrees of freedom to
reliably solve for both amplitude and phase distortions.

Table~\ref{tab:errors} summarizes a set of numerical experiments that
illustrate these criteria.
Each experiment randomly generates a ground-truth set of mismatches
$\hat\gamma_p$, $\hat\tau_p$, and $\hat\alpha_p$, and attempts to
recover them by applying equations~\eqref{eq:Gamma_sol}
and~\eqref{eq:alpha_sol} to synthetic sampled data.
The table reports reconstruction root-mean-square (RMS) errors,
$\RMS(\hat\gamma_p - \gamma_p)$, $\RMS(\hat\tau_p - \tau_p)$, and
$\RMS(\hat\alpha_p - \alpha_p)$, under different configurations of $P$
and $N_\delta$.

The results are color-coded to show the sufficiency of the sampling
configuration:
\begin{itemize}
\item {\color{MidnightBlue}Blue} entries indicate that both $N_\delta
  > P$ and $N_T N_\delta \ge 3P$ are satisfied, meeting the
  theoretical minimum for solving all mismatch parameters (phase,
  gain, and offset).
\item {\color{OliveGreen}Green} entries mark cases where $N_T N_\delta
  \ge 2P$ but $N_\delta \le P$, satisfying only the condition for
  solving the offset mismatches $\alpha_p$.
\end{itemize}

Determining the minimal $N_\delta$ (samples per period) depends on the
relationship between $P$ and $N_\delta$.
However, a particularly elegant and efficient configuration occurs
when $N_\delta = P + 1$.
In this case, because $P$ and $P+1$ are \emph{coprime}, we have
\begin{align}
  N_T N_\delta = \LCM(P, N_\delta) = P(P+1),
\end{align}
which guarantees that each ADC samples the reference signal $P + 1$
times, evenly distributed across $P$ periods.
This yields exactly $P(P+1) \ge 3P$ samples, satisfying all
requirements to solve for gain, delay, and offset mismatches.

\section{Convergence Properties}
\label{sec:noise}

\begin{figure*}[t]
  \centering \includegraphics[width=0.5\textwidth]{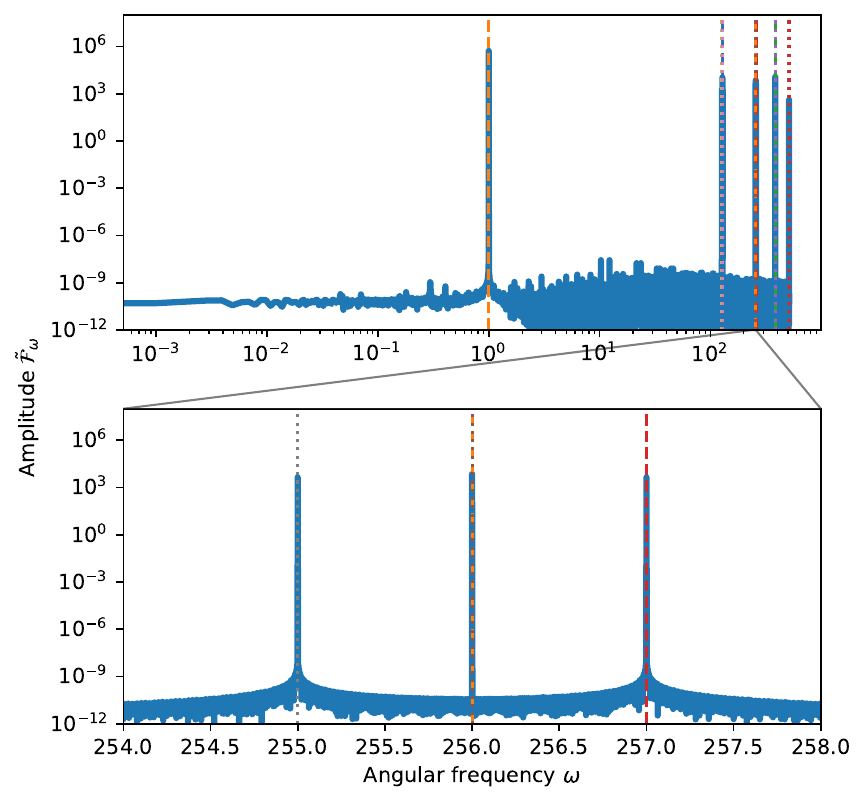}%
  \includegraphics[width=0.5\textwidth]{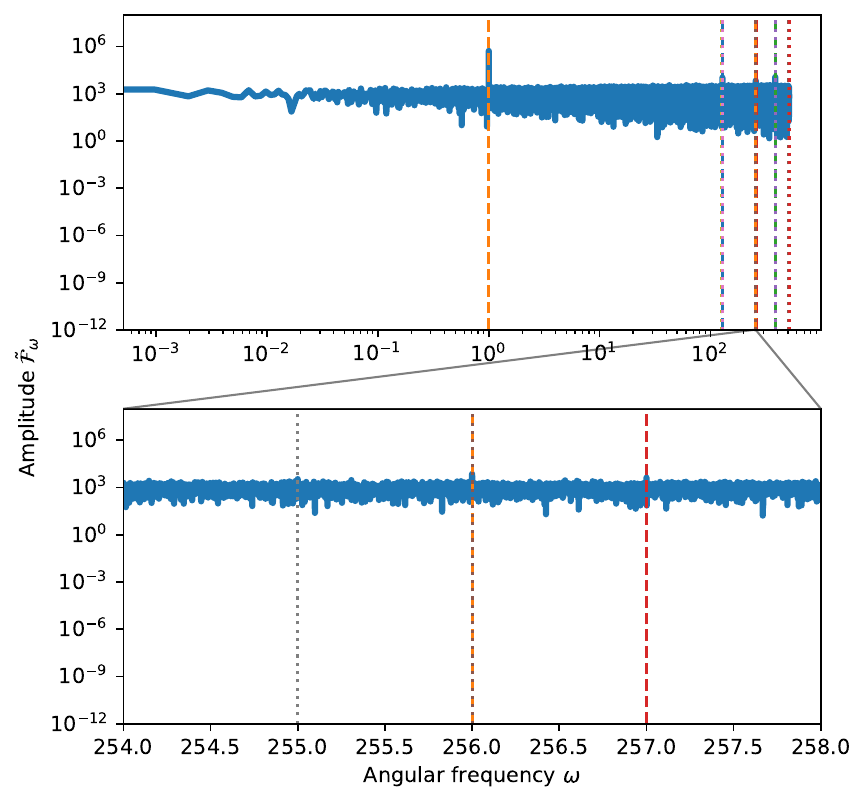}
  \caption{%
    (\emph{Left}) Frequency spectrum of the signal before applying the
    proposed direct calibration algorithm, illustrating the effects of
    phase delay, gain, and offset mismatches in a 4-ADC interleaving system.
    Dashed and dotted vertical lines in different colors mark the
    frequencies at $\Omega + m \Omega_\Delta$ (see
    equation~\eqref{eq:classI}) and $m' \Omega_\Delta$ (see
    equation~\eqref{eq:classII}), respectively.
    The top panel shows the full spectrum across a broad range of
    angular frequencies, while the bottom panel provides a zoomed-in
    view of a specific band, clearly revealing spurious peaks
    primarily caused by phase delay mismatches.
    (\emph{Right}) Same as the left, but with additive noise.
    The noise level is comparable to that of the reference signal.
    Although the spurious peaks remain visible, their amplitudes are
    partially masked by the noise.%
  }
  \label{fig:spectra}
\end{figure*}

\begin{figure}
  \centering
  \includegraphics[width=\columnwidth]{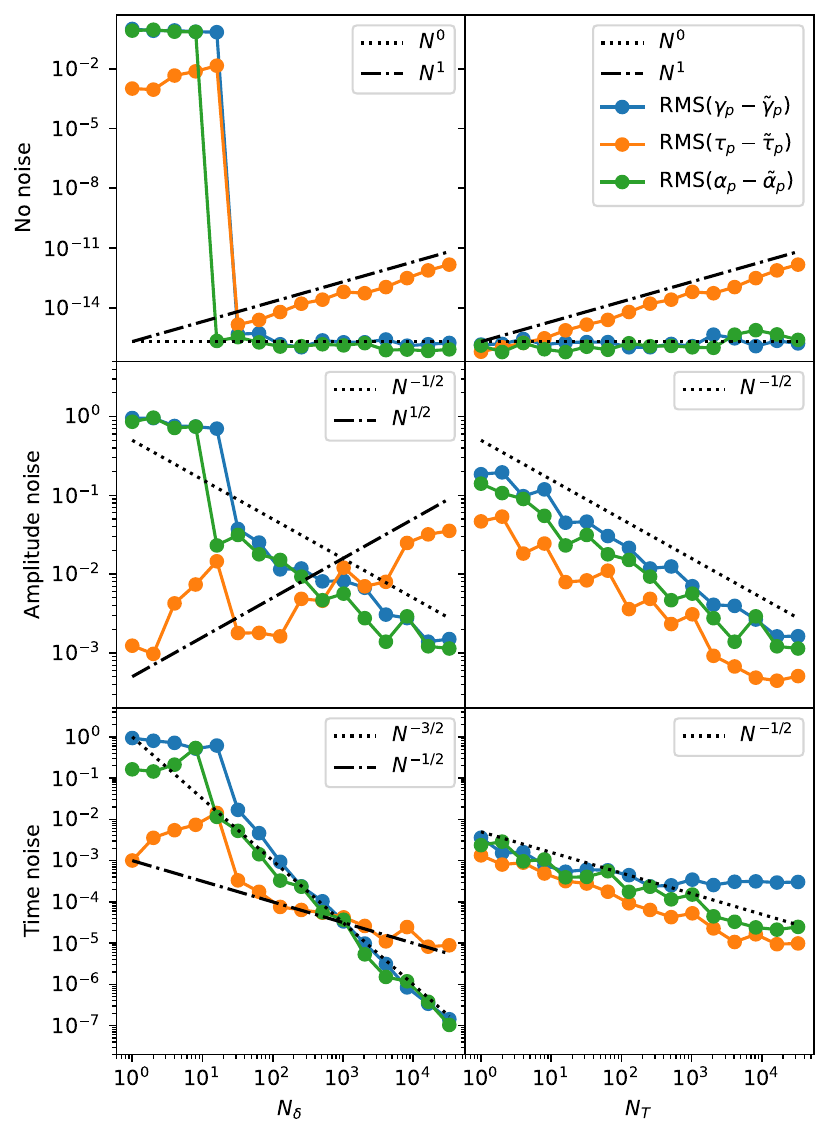}
  \caption{%
    Convergence properties of the proposed direct calibration
    algorithm as a function of the number of samples.
    (\emph{Left column}) The number of periods is fixed at $N_T =
    256$, while the number of samples per period $N_\delta$ is varied.
    (\emph{Right column}) The number of samples per period is fixed at
    $N_\delta = 256$, while the number of periods $N_T$ is varied.
    Colored circles indicate the RMS errors in recovering each
    mismatch parameter: phase delay $\tau_p$, gain $\gamma_p$, and
    offset $\alpha_p$.
    The dotted and dash-dotted lines show reference slopes for ideal
    scaling behavior, allowing comparison to theoretical convergence
    rates.
    These plots demonstrate that different mismatch parameters exhibit
    distinct but predicted convergence characteristics.
    See section~\ref{sec:noise} for details.%
  }
  \label{fig:convergence}
\end{figure}

When using the minimal number of samples required for mismatch
recovery, any noise in the system can significantly degrade the
accuracy of the solution.
Figure~\ref{fig:spectra} illustrates how phase delay, gain, and offset
mismatches, as well as noise, affect the frequency spectrum of the
sampled signal before calibration.
The left panels show spurious peaks appear at frequency offsets of
$\Omega + m\Omega_\Delta$ as a result of phase delay mismatches.
The right panels show, in the presence of noise, these features remain
visible but are partially masked, motivating the need for oversampling
strategies to enhance the signal-to-noise ratio (SNR) in the
calibration process.

To mitigate noise, it is essential to oversample the reference signal
so that noise effects can be averaged out.
We consider two main strategies to increase sampling redundancy.
First, the number of samples per period, $N_\delta$, can be increased
to oversample the reference waveform in time.
For a physical \PADC system where $P$ and $\Delta$ are fixed, this can
be achieved by tuning the reference signal's frequency $\Omega =
2\pi/T$ so that its period satisfies $T = N_\delta \delta$.
It is also important to note that real ADCs have finite integration
times.
Increasing $T$ helps ensure that this finite integration time remains
negligible in the calibration procedure.
Second, the number of sampled periods, $N_T$, can be increased while
keeping $N_\delta$ fixed.
This allows for temporal averaging across many cycles of the reference
signal.

In our numerical experiments shown in Table~\ref{tab:errors} and
Figure~\ref{fig:convergence}, we fix the interleaved sampling interval
$\delta$ to represent a hardware constraint.
The actual value is not important, but for completeness, our numerical
experiment uses $\delta = 2\pi/1024$.
The number of interleaved ADCs, $P$, determines the sampling interval
for each ADC.
We then vary $N_\delta$ and $N_T$ to investigate how recovery accuracy
depends on total sampling.

Figure~\ref{fig:convergence} quantifies the convergence behavior of
our calibration algorithm.
In the left column, we fix $N_T = 256$ and vary $N_\delta$; in the
right column, we fix $N_\delta = 256$ and vary $N_T$.
Each panel reports the RMS error in recovering the mismatch
parameters: phase delay $\tau_p$, gain $\gamma_p$, and offset
$\alpha_p$, under three different noise conditions:

\textbf{Top row (no noise):}
When $N_\delta \lesssim 200$, the number of samples is below the
theoretical minimum, and all recovered parameters exhibit significant
errors.
Once $N_\delta \gtrsim 200$, the error in $\gamma_p$ and $\alpha_p$
drops to machine precision, consistent with the deterministic
predictions of our theory.
However, even in this regime, the phase delay $\tau_p$ continues to
scale linearly with $N_\delta$.
This is explained by the fact that increasing $N_\delta$ lowers the
reference frequency $\Omega$, and since $\tau_p =
\arg(\varGamma_p)/\Omega$, small phase errors in $\varGamma_p$ get
amplified.
When varying $N_T$ at fixed $N_\delta$, the errors in $\gamma_p$ and
$\alpha_p$ remain flat, while $\tau_p$ still exhibits increasing error
due to its dependence on total integration time.

\textbf{Middle row (additive amplitude noise):}
Errors in $\gamma_p$ and $\alpha_p$ scale approximately as
$1/\sqrt{N_\delta}$ and $1/\sqrt{N_T}$, reflecting classical
statistical averaging.
However, for $\tau_p$, the scaling with $N_\delta$ becomes worse due
to the additional $\Omega^{-1} \propto N_\delta$ factor, resulting in
$\tau_p$ errors scaling roughly as $\sqrt{N_\delta}$.
When $N_T$ increases (right column), all mismatch parameters benefit
equally from temporal averaging and scale as $1/\sqrt{N_T}$.

\textbf{Bottom row (sampling jitter):}
When noise affects the sampling times directly (timing jitter), the
error in $\tau_p$ dominates and decreases approximately as
$1/\sqrt{N_\delta}$.
Interestingly, errors in $\gamma_p$ and $\alpha_p$ decrease faster,
with scaling close to $1/N_\delta^{3/2}$.
This behavior likely arises because jitter impacts phase recovery most
directly, while the amplitude and offset terms average out more
efficiently.

These results suggest that while gain and offset mismatches can be
reliably recovered at relatively low SNRs, phase delay recovery
requires higher sampling density and cleaner reference signals to
achieve comparable accuracy.

In all cases, the constants of proportionality in the error scaling
depend on the amplitude and nature of the noise.
In physical systems, such noise is determined by factors such as
thermal fluctuations, system temperature, and clock phase jitter.
As a practical guideline, we recommend performing a system-specific
calibration sensitivity analysis to determine the minimal values of
$N_\delta$ and $N_T$ required to meet the desired accuracy thresholds
for each mismatch parameter.
This approach balances hardware constraints and calibration precision,
and ensures robust operation of interleaved ADC systems in noisy
environments.

\section{A Direct Calibration Algorithm}
\label{sec:algo}

We now present an efficient direct calibration algorithm to recover
gain, phase delay, and offset mismatches ($\gamma_p$, $\tau_p$, and
$\alpha_p$) in a \PADC interleaving system with computation complexity
$\mathcal{O}(N \min(\log N, P))$.

The hardware configuration specifies the number of interleaving ADCs
$P$ and their sampling interval $\Delta$, which determines the
interleaved sampling interval $\delta = \Delta / P$.
To perform calibration, we sample a known sinusoidal reference signal
with angular frequency $\Omega$ at period $T = N_\delta \delta$.
Here, $N_\delta$ is the number of samples per period across all ADCs,
and $N_T$ is the number of full signal periods sampled.
This results in a total of $N = N_\delta N_T$ sampled points.

Our goal is to extract the Fourier coefficients
$\tilde{\mathcal{F}}_{\Omega + m\Omega_\Delta}$ and
$\tilde{\mathcal{F}}_{m\Omega_\Delta}$ from the sampled data
$\tilde{f}_n$, where $\Omega_\Delta = 2\pi/\Delta$.
These coefficients are then used to compute $\tilde{\mathcal{R}}_m$
and $\tilde{\mathcal{A}}_m$, from which the complex gains $\varGamma_p
= \gamma_p e^{i\Omega \tau_p}$ and offsets $\alpha_p$ are recovered
via inverse discrete Fourier transforms, as given in
equations~\eqref{eq:Gamma_sol} and~\eqref{eq:alpha_sol}.

To compute the Fourier transform of the sampled signal, one may use a
Fast Fourier Transform (FFT) or a direct (matrix-based) Discrete
Fourier Transform (DFT), depending on the problem size.
If the total number of samples $N$ is large ($N \gg P$), FFT provides
a faster solution.
For small-scale problems with modest $P$, DFT may be efficient and
more flexible for selecting arbitrary frequencies.
Once $\varGamma_p$ and $\alpha_p$ are computed, we extract the gain
and phase mismatches using $\gamma_p = |\varGamma_p|$ and $\tau_p =
\arg(\varGamma_p)/\Omega$.

The entire procedure is summarized in Algorithm~\ref{alg:cal}.
Once the mismatch parameters are recovered, corrections can be applied
in hardware by adjusting ADC clocks or digital compensation blocks, or
in software during post-processing if the mismatches are small and
stable.
Because the algorithm is fast and lightweight, it can be run
repeatedly to track and correct for slow mismatch drifts during
operation, enabling robust, ongoing calibration with minimal overhead.

\begin{algorithm}[h]
  \caption{The proposed direct Fourier-based calibration algorithm for
    recovering gain, phase delay, and offset mismatches in interleaved
    ADC systems.
    Note that $m = 0, 1, \dots, P-1$.}
  \label{alg:cal}

  \KwResult{$\alpha_p$, $\gamma_p$, and $\tau_p$ for $p = 0, 1, \dots, P-1$}

  \While{need calibration}{
    $\tilde{f}_n \gets N$ samples from reference signal $\cos(\Omega t)$\;
    \eIf{$2 P > 5 \log N$}{
      $\tilde{\mathcal{F}}_\omega \gets \mathrm{FFT} \tilde{f}_n$\;
    }{
      $\tilde{\mathcal{F}}_\omega \gets \mathrm{DFT} \tilde{f}_n$ for $\omega = \Omega + m\Omega_\Delta$ or $m\Omega_\Delta$\;
    }
    $\tilde{\mathcal{R}}_m \gets \tilde{\mathcal{F}}_{\Omega + m\Omega_\Delta} / (A/2)$\;
    $\tilde{\mathcal{A}}_m \gets \tilde{\mathcal{F}}_{m\Omega_\Delta}$\;
    \eIf{$P \gg 5 \log P$}{
      $\varGamma_p \gets \mathrm{FFT}^{-1} \tilde{\mathcal{R}}_m$\;
      $\alpha_p    \gets \mathrm{FFT}^{-1} \tilde{\mathcal{A}}_m$\;
    }{
      $\varGamma_p \gets \mathrm{DFT}^{-1} \tilde{\mathcal{R}}_m$\;
      $\alpha_p    \gets \mathrm{DFT}^{-1} \tilde{\mathcal{A}}_m$\;
    }
    $\gamma_p    \gets |\varGamma_p|$\;
    $\tau_p      \gets \arg(\varGamma_p)/\Omega$\;
    adjust hardware and/or post-process parameters according to $\tau_p$, $\gamma_p$, and $\alpha_p$\;
  }
\end{algorithm}

\section{Discussion and Conclusion}
\label{sec:discussions}

In this paper, we have presented a novel direct calibration algorithm
to correct phase delay, gain, and offset mismatches in ADC
interleaving systems.
The algorithm is based on a deterministic analysis using Fourier
transforms of a sinusoidal reference signal, enabling efficient and
accurate extraction of mismatch parameters from the frequency-domain
representation of the sampled data.
By avoiding iterative search or optimization, our method achieves both
computational efficiency and mathematical transparency.

We have demonstrated that the mismatch parameters can be recovered
from the observed signal.
This enables flexible implementation strategies: the mismatch
corrections can be applied in \emph{hardware}, through adjustments to
ADC sampling clocks, gain amplifiers, or DC offset registers; in
\emph{software}, by compensating for the recovered mismatches during
post-processing; or in a \emph{hybrid} approach, where coarse
corrections are handled in hardware and fine-tuning is performed in
software post-processing.
Because our algorithm is fast and lightweight, it can be invoked
repeatedly to track slow drift in mismatch parameters, which is
important for long-duration observations or thermally varying
environments.

We validated the algorithm through a series of numerical experiments
that demonstrate the minimal sampling requirements for robust
calibration and the convergence behavior under different noise models.
In particular, we showed that phase delay mismatch recovery is more
sensitive to sampling density and noise compared to gain and offset
mismatches in our setup.

The proposed approach is well-suited for high-speed, high-precision
applications such as radio astronomy and VLBI.
VLBI systems often rely on tightly synchronized, interleaved ADC
architectures to achieve extremely high bandwidths.
Spurious tones and spectral leakage from uncorrected mismatches can
compromise image reconstruction and astrometric precision.
Recent work by the EHT collaboration
\citep[e.g.,][]{2022ApJ...930L..13E} has emphasized the need for
calibration strategies that scale with increasing ADC count and
bandwidth.
Our method provides a deterministic and scalable solution to this
challenge.
By enabling reliable calibration with arbitrary numbers of interleaved
ADCs, it opens a pathway to even broader bandwidths and finer time
resolution in next-generation instruments.

Several promising avenues exist for extending this work.
First, implementing the algorithm in FPGA or real-time DSP systems
will allow on-the-fly mismatch correction, enabling real-time
calibration in operational environments.
Second, incorporating feedback mechanisms to track slow drifts in
mismatch parameters (e.g., due to thermal changes) could allow the
system to maintain calibration autonomously over time.
Third, extending the method to handle non-sinusoidal or multi-tone
calibration signals could facilitate mismatch recovery without
interrupting the science data stream.
Finally, studying how ADC quantization influences mismatch recovery
will be crucial for low-bit-depth systems.
In particular, we propose analyzing how quantization and noise levels
jointly limit the maximum phase mismatch that can be reliably
corrected in software.

The direct calibration algorithm presented in this work offers an
efficient, deterministic, and scalable solution to correct ADC
interleaving mismatches.
Its low computational overhead, compatibility with hardware and
software workflows, and high accuracy make it an attractive approach
for future high-throughput data acquisition systems.
In the context of radio astronomy and VLBI, it contributes to
next-generation calibration pipelines, allowing for wider bandwidths,
improved signal fidelity, and enhanced scientific capabilities.


\bibliographystyle{aasjournal}
\bibliography{refs,main}

\end{document}